# State-aware protein-ligand complex prediction using AlphaFold3 with purified sequences

Enming Xing[a], Junjie Zhang[a], Shen Wang[a], Xiaolin Cheng [a,b*]

[a] Division of Medicinal Chemistry and Pharmacognosy, College of Pharmacy, The Ohio State University, Columbus OH, 43210, USA

[b] Translational Data Analytics Institute, The Ohio State University, Columbus, OH 43210, USA

*Corresponding Authors: Xiaolin Cheng (cheng.1302@osu.edu)*

**Abstract**

Deep learning–based prediction of protein–ligand complexes has advanced significantly with the development of architectures such as AlphaFold3, Boltz-1, Chai-1, Protenix, and NeuralPlexer. Multiple sequence alignment (MSA) has been a key input, providing coevolutionary information critical for structural inference. However, recent benchmarks reveal a major limitation: these models often memorize ligand poses from training data and perform poorly on novel chemotypes or dynamic binding events involving substantial conformational changes in binding pockets. To overcome this, we introduced a state-aware protein–ligand prediction strategy leveraging purified sequence subsets generated by AF-ClaSeq—a method previously developed by our group. AF-ClaSeq isolates coevolutionary signals and selects sequences that preferentially encode distinct structural states as predicted by AlphaFold2. By applying MSA-derived conformational restraints, we observed significant improvements in predicting ligand poses. In cases where AlphaFold3 previously failed—producing incorrect ligand placements and associated protein conformations—we were able to correct the predictions by using sequence subsets corresponding to the relevant functional state, such as the inactive form of an enzyme bound to a negative allosteric modulator. We believe this approach represents a powerful and generalizable strategy for improving protein–ligand complex predictions, with potential applications across a broad range of molecular modeling tasks.

**Introduction:**

Protein–ligand interactions underlie almost all aspects of chemical biology and drug discovery, making accurate prediction of protein–ligand complex structures a long-sought goal in computational biology[1]. Classical *in silico* approaches, such as molecular docking, have been indispensable for virtual screening and pose prediction, but they face well-known limitations. Rigid-receptor docking and simplistic scoring functions often fail to capture the intricacies of binding poses and protein flexibility, and cannot distinguish the false positive entries[2]. These challenges—compounded by the limited availability of high-quality protein–ligand complex structures for model training—have historically hindered reliable ligand pose prediction, especially in cases requiring induced-fit conformational changes.

In recent years, deep learning-based methods have revolutionized protein or protein-ligand structure prediction. AlphaFold2 (AF2) delivered unprecedented accuracy in predicting protein structures from sequence, effectively solving the 50-year protein folding problem for many targets[3]. Its transformative success stems from large multiple sequence alignments (MSAs) and coevolutionary signals, which allowed a neural network (the Evoformer) to infer inter-residue pairwise contacts and orientations with atomic-level precision. Furthermore, AlphaFold-Multimer[4] adapted the model to protein–protein complexes by pairing MSAs of interacting partners, enabling accurate quaternary structure predictions from sequence alone. Likewise, RoseTTAFold[5] introduced a multi-track network to jointly encode protein pairs, presaging the extension of single-chain folding networks to higher-order complexes. These early developments proved that prediction of intermolecular interfaces greatly benefit from coevolutionary information encoded in MSA, which also paved the way for protein-ligand complex scenario. AlphaFold 3 (AF3) exemplifies this progress: it introduced a substantially updated, diffusion-based architecture capable of predicting the *joint* 3D structure of protein complexes that include not only other proteins but also nucleic acids, small molecules, ions, and post-translational modifications[1]. By unifying nearly all major interaction types within a unified deep learning framework, AlphaFold3 achieved higher accuracy on protein–ligand complexes than traditional docking methods, and likewise outperformed specialized predictors in protein–RNA/DNA and antibody–antigen modeling. Notably, AF3's ligand pose predictions on challenging benchmarks (e.g. the PoseBusters set[6]) rival the best available methods: the model achieves successful pose predictions for roughly three-quarters of cases, substantially surpassing classical docking programs. Several AlphaFold3-inspired systems also emerged – notably Protenix[7], Boltz-1[8], and Chai-1[9] – which reproduce or enhance AF3's capabilities. Protenix, for example, is an open-source reimplementation of AF3 that provides full training code and model weights, and it matches or even slightly outperforms the original AF3 on benchmarks of protein–ligand binding (validated on PoseBusters) as well as protein–protein and protein–DNA/RNA complexes. Similarly, Boltz-1 achieves AlphaFold3-level accuracy for 3D biomolecular interactions and introduces innovations like pocket-conditioned inference to better handle ligand placement. Chai-1, a multi-modal foundation model, attains comparable performance to AF3 (77% success on PoseBusters vs. 76% for AF3) while offering practical advantages such as an open commercial license and the ability to generate accurate structures *without* an MSA. These tools, along with AlphaFold3 itself, represent the state-of-the-art in protein–ligand complex structure prediction. They leverage deep learning architectures (from attention-based transformers to diffusion generative models) trained on the wealth of structural data in the Protein Data Bank, and they have effectively brought AI-driven protein–ligand co-folding to the fore of computational chemistry.

Despite this remarkable progress, important challenges remain for current protein–ligand prediction models. Firstly, their handling of novel ligands and chemotypes that fall outside the distribution of training data. While models like AF3 and its derivatives generalize impressively, their accuracy can deteriorate for ligands with very unusual scaffolds, in part because the neural networks may not have learned features for entirely new functional groups or stereochemistry. Indeed, recent assessments suggest that high prediction success often correlates with the query protein–ligand system having close analogs in the training set, raising concerns that some methods may rely on memorized interactions[10, 11]. As a result, genuinely novel drug-like molecules or those engaging unique binding pockets can still confound even the best algorithms. Secondly, a major challenge lies in the conformational plasticity of proteins. Most current models predict only a single bound conformation, effectively assuming the binding pocket is pre-formed. In reality, many proteins undergo substantial induced-fit changes or allosteric rearrangements when binding to a ligand. Standard AlphaFold-based pipelines, which input a fixed sequence (and MSA) and output one predominant structure, may miss alternative states that are crucial for accommodating certain ligands. AlphaFold2, for instance, tends to predict a static ground-state structure and cannot easily switch to a radically different conformation without additional guidance[12]. Some state-of-the-art methods have begun to address this: NeuralPLexer[13], a deep generative approach, directly tackles flexible binding by *sampling* alternative protein–ligand conformations. Using a diffusion model with integrated biophysical constraints, NeuralPLexer can generate both the ligand-free (apo) and ligand-bound states of a protein in a single framework. This state-specific modeling enabled NeuralPLexer to achieve excellent results on blind docking benchmarks that require receptor flexibility, and it consistently outperformed AlphaFold2 on targets known to undergo large conformational changes upon ligand binding. Nonetheless, even NeuralPLexer and AlphaFold3 have practical limits on the magnitude of conformational rearrangement they can reliably capture, and they may struggle with cases where subtle sequence variations or allosteric effectors determine the binding mode[12]. Another crucial factor influencing models' output is the use of evolutionary information. The reliance on multiple sequence alignments (MSAs) and coevolutionary signals was essential to AlphaFold's dramatic improvement in accuracy. However, noisy or heterogeneous evolutionary signals can mislead the model, often causing it to favor the predominant structural state encoded in the MSA while ignoring subtle signals associated with less common conformations. Given this inherent heterogeneity, attention-based transformer networks struggle to disentangle overlapping co-evolutionary signals linked to different structural states. As a result, these signals may be averaged or obscured, confounding the model's ability to make accurate and nuanced structural predictions.

Motivated by these observations, we developed a state-aware protein–ligand prediction strategy that leverages evolutionary sub-signals to improve protein-ligand complex prediction accuracy. Our previously published framework, **AF-ClaSeq**[14], isolates subtle co-evolutionary signals from MSAs through iterative enrichment, statistical bootstrapping, and voting mechanisms. By extracting subsets of sequences that preferentially encode distinct structural states, AF-ClaSeq enables confident prediction of alternative conformations. Our findings reveal that the successful sampling of alternative states depends not on MSA depth but on sequence purity. Intriguingly, purified sequences encoding specific structural states are distributed across phylogenetic clades and superfamilies, rather than confined to specific lineages. AF-ClaSeq thus extends AlphaFold2's capabilities by uncovering hidden structural plasticity essential for dynamic protein function and drug design.

In this work, we further expand the utility of purified sequences by introducing a state-aware protein–ligand prediction approach that integrates AF-ClaSeq derived sequence subsets into the structure prediction pipeline. Given that AlphaFold3 and AlphaFold2 utilize similar MSA processing mechanisms through attention-based architectures, differing primarily in AF3's ability to incorporate additional molecular inputs, we expect that purified sequences for specific state predictions in AF2 can be effectively transferred to AF3. By using AF-ClaSeq to systematically purify MSAs and enrich coevolutionary signals specific to the specific ligand-bound and functionally relevant conformations, our method guides co-folding algorithms toward conformational states that better accommodate the ligand. This strategy addresses key challenges in modeling novel chemotypes and induced-fit effect by enforcing a conformational constraint encoded in the purified MSAs. We show that incorporating AF-ClaSeq-derived sequence subsets into AF3 framework leads to more accurate ligand placement and pocket geometries than its default results. Collectively, our approach establishes a powerful framework for ligand pose prediction that is both evolutionarily informed and conformation-sensitive, enabling more precise modeling of protein–ligand interactions in complex, dynamic systems.

**Results**

**Sequence Purification of Epidermal Growth Factor Receptor (EGFR) Inactive State Improves Ligand Pose Prediction for Allosteric Inhibitors**

As a critical kinase family member, epidermal growth factor receptor (EGFR) has been an attractive drug target in oncology[15, 16]. However, the mutations leading to drug resistance are very prone to occur, therefore new generation EGFR inhibitors or allosteric modulators are always demanding in multiple disease areas such as non-small cell lung cancer (NSCLC)[17, 18]. So far, there have been 351 deposited structures of human EGFR protein (under UniProt accession code P00533). The vast majority of human EGFR kinase–ligand complexes in the PDB are ATP-competitive Type I inhibitors (e.g., gefitinib, erlotinib), comprising well over 80% of entries. A smaller yet significant fraction (~10%) are irreversible Type IV covalent inhibitors (e.g., afatinib[19], osimertinib[20]) that form a bond with Cys797. Type I½ binders—ATP-site ligands preferring a DFG-in but "inactive" conformation (e.g., lapatinib[21])—account for a few percent of structures. Classical Type II DFG-out inhibitors are very rare for EGFR (<1%), and cases of Type III allosteric inhibitors (e.g., EAI045[22], JBJ-04-125-02[23]) are also few, appearing only in specialized mutant-selective complexes (<10%). The scarcity of allosteric inhibitor structures in training datasets poses significant challenges for AlphaFold3 (AF3) predictions. For example, Obst-Sander et al. from Roche published a series of allosteric EGFR L858R inhibitors for non-small cell lung cancer treatment[24], depositing four PDB structures with small molecule compounds co-crystallized with the EGFR protein (PDB codes: 8A27, 8A2A, 8A2B, and 8A2D). Remarkably, we found that under default settings, AF3 could only correctly predict one of these structures (PDB code 8A27), while failing to accurately predict the other three complexes (**Figure 1**). The predictions showed errors not only in the protein conformation but also in the ligand binding pose, with ligand RMSD ranging from 14.9 Å to 19.1 Å after global alignment. Based on published benchmarking studies[10], AF3 tends to predict ligand interactions that resemble compounds or interaction modes frequently present in its training dataset, leading to a bias toward well-represented binding modes. This prediction bias poses a fundamental challenge for AlphaFold3 in drug design contexts, where the imperative to develop novel proprietary chemical entities with novel mechanisms of action often conflicts with the model's tendency to predict familiar interaction patterns. Therefore, enabling AF3 to reliably predict new chemotypes with unseen binding modes is critical for the application of computational drug design. The problem is especially acute for allosteric inhibitors, which bind at sites remote from the traditional ATP pocket and may not be well represented in existing structural databases.

To address this challenge, we hypothesized that enforcing a conformational constraint through input MSA could improve the protein-ligand complex prediction. Since incorrectly predicted structures show EGFR in the active state when bound to allosteric inhibitors—contrary to the expected inactive state—we sought to purify sequences that bias toward the inactive conformation. We adopted the same collective variables used by Shan et al. to define EGFR conformational states, designating PDB structure 2ITP as the active (DFG-in) reference and PDB structure 2GS7 as the inactive (Src-like inactive) reference[25] (**Figure 2A**). RMSD calculations focused on the Cα atoms of the αC helix and activation loop (A-Loop) two-turn helix regions (residues 756–769 and 857–863).

Starting from a DeepMSA2-generated MSA containing 49,743 sequences, we applied a coverage threshold of 0.6, yielding 43,365 filtered sequences. To inspect the initial conformational distribution encoded within the entire pool, we divided them into groups of 28 sequences and shuffled randomly 10 times, performing a total of 15,490 predictions. The initial distribution analysis revealed that the majority of conformational information encoded in the MSA was biased toward the active state rather than the inactive state, explaining why default AF3 predictions favor the active state and fail to yield correct allosteric binding poses with an inactive state. Therefore, our first step was to enrich the sequence distribution toward the inactive state before M-fold sampling, as direct sampling would be computationally prohibitive. An iterative enrichment was performed, dividing sequences into groups of 6 and shuffling randomly 10 times per iteration. At each iteration, we selected the sequences with lowest 15% of αC helix and activation loop two-turn helix regions RMSD with respect to the inactive state (PDB structure 2GS7) for the next iteration. After four iterations, we observed that the number of predicted structures with lower RMSD begin to emerge compared to the first iteration of shuffling.

The sequence pool from iteration 4 was then used for M-fold sampling, where 1,405 sequences were divided into 233 groups of 6 sequences each, resulting in 233 prediction runs, with all structures plotted as shown in **Figure 2B**. The scatter plot was color-coded in two different ways: one using the global pLDDT score by AF2, and the other using the local pLDDT of residues 756–769 and 857–863 (αC helix and activation loop two-turn helix regions). Since these two regions mainly comprise loops and highly dynamic helical secondary structures, the local pLDDT is below the global average. However, we can still observe that when structures approach either state endpoint—active or inactive—the local pLDDT appears higher than those structures distant from these two states. Based on the two RMSD metrics, we were able to vote for sequences that contribute most to each state using the method we previously described (**Figure 2C**). We employed two levels of voting: in the first level, without enforcement of any threshold, any sequence that showed more occurrence in one bin than the others were voted for this bin. In the second level, sequences were voted not only when their occurrence was the most frequent but also exceeded 0.15. After enforcing the threshold in voting, we did not obtain many sequences for the active state but acquired very few sequences for the inactive state. These sequences are considered highly biased toward the inactive state.

AF2 predictions of the protein alone were performed using the purified sequences. When compared to randomly selected sequences of the same number, we found that the purified sequences showed high convergence in the prediction distribution toward one conformational state (**Figure 2D, E**). These purified sequences corresponding to the inactive state were then used for AF3 predictions, where the customized input MSA was provided along with the SMILES of the ligand. No templates were used as input for the predictions, relying solely on the MSA compiled from purified sequences (**Figure 3, 4**). Using ten random seeds with five structures generated per seed, a total of 50 structures were produced. As expected, the default prediction yielded very divergent and inconsistent prediction results of ligand poses, which also showed low ligand atomic pLDDT scores. When using sequences purified from normal voting without enforced thresholds (bin 7 and 8 sequences), though the cases of 8A2B and 8A2D performed relatively better than 8A2A, none showed 100% successful prediction. However, when using a subset of sequences purified based on voting with enforced thresholds (bin 9 sequences), the prediction results were astonishingly good—all showed low ligand RMSD values near or below 2.5 Å, with elevated ligand average atomic pLDDT scores and high consistency.

Interestingly, when using the purified inactive state sequences for prediction in AF2, it appears that bin 7 and 8 sequences produced better prediction results than bin 9, which may indicate differences in how AF2 and AF3 transform MSA-encoded information into structural records. However, comparing the AF3 prediction results from bin 7, 8, and bin 9 sequences, it appears that sequences with strong bias toward a specific state benefit the convergence of prediction results. The ligand view of predictions using bin 9 purified sequences is shown in **Figure 4**, where predictions using purified sequences are colored in green compared to the experimental structure from the PDB.

**Sequence purification of ligand bound interleukin-1β (IL-1β) loop conformation corrects IL-1β/ligand complex prediction**

To further validate the value of conformational restraints in improving AF3 predictions, we deployed it on another case, interleukin-1β (IL-1β), a critical pro-inflammatory cytokine[26, 27] where conformational changes in the β4-5 and β7-8 loops are essential for allosteric inhibitor binding. Recent structural studies by Hommel et al. revealed that the small molecule antagonist (S)-2 binds to a cryptic pocket formed by displacement of the β4-5 loop by up to 11 Å from its position in the mature cytokine, creating an allosteric binding site that prevents proper IL-1β/IL-1R1 interaction[28]. However, default AlphaFold3 predictions consistently failed to capture this ligand-induced conformational state, instead predicting the conventional IL-1β structure where the β4-5 loop remains in its native position, unable to accommodate the cryptic pocket formation required for allosteric inhibitor binding (**Figure 5A**). Interestingly, default AF2 predictions of the protein alone also produced a similar conformation where the β4-5 loop was not in the cryptic pocket state (**Figure 5B**). Starting with 4,129 sequences obtained from DeepMSA2, we performed a similar iterative enrichment approach as used for EGFR, where RMSD relative to the β4-5 loop (residues 46-55) and β7-8 loop (residues 86-96) was used as the metric to enrich sequences that produced structures with low RMSD values. This resulted in 1,476 sequences as an enriched set. These sequences were used for M-fold sampling, and a focused range of PDB structures with RMSD relative to the β4-5 loop lower than 2.5 Å and β7-8 loop lower than 3.0 Å was selected (**Figure 5C**). The corresponding MSAs were located for each predicted PDB file, and the sequences within these MSAs were sorted with statistical ranking performed to identify which sequences appeared most frequently in this region, indicating significant contribution to this conformation. We identified the top 10 and top 20 sequences that are the most frequent occur in the predicted structures in this range and performed AF3 predictions with MSAs compiled by these frequently occurring sequences as input. No templates were searched, and the SMILES of the ligand was input alongside the MSA.

As shown in **Figure 6**, the default prediction failed to predict both the ligand pose and protein pose correctly, illustrated by the structure model visualization shown in purple in the figure. The top 10 most frequent sequences showed much better results with only very few outliers, and most predicted models displayed extremely low RMSD values (below 0.5 Å). With the top 20 sequences, all 40 structures were predicted to align perfectly with the experimental structure, showing very strong convergence and conformational constraint on the prediction outcome. This is consistent with the conclusions we reached in the EGFR case.

**Conclusion**

In this study, we have demonstrated that sequence purification provides a powerful solution to overcome AlphaFold3's inherent limitation in predicting novel small molecular ligand in rare interaction patterns. Through comprehensive analysis of two distinct allosteric inhibitor systems—EGFR L858R mutant with fourth-generation inhibitors and IL-1β with cryptic pocket antagonists—we showed that default AF3 predictions consistently fail to capture the conformational states required for accurate allosteric drug-target interactions or cryptic pocket formation, yet prone to memorize those similar entries represented in training data. Our iterative sequence enrichment approach, guided by conformational state-specific RMSD metrics rather than homologous clustering, successfully identified purified MSA subsets that bias predictions toward functionally relevant inactive or allosteric conformations. The dramatic improvements achieved—from complete prediction failures to sub-2.5 Å ligand RMSD accuracy with high pLDDT scores and structural convergence—demonstrate the critical importance of conformational constraints in computational drug discovery. A key advantage of our approach is that the sequence purification process operates independently of specific target structure

information—it does not require prior knowledge of individual inhibitor-bound structures (such as PDB codes 8A2A, 8A2B, 8A2D, or 8C3U). Instead, the methodology relies on generalized conformational state references (such as active 2ITP and inactive 2GS7 states for EGFR, or apo versus allosterically bound states for IL-1β), making it broadly applicable across diverse protein families and inhibitor classes. This universality means that whenever an allosteric inhibitor is expected to stabilize alternative conformations, or when researchers have solved an initial crystal structure and wish to predict binding of similar ligand analogs, our method can provide a generalizable framework for conformational constraint-guided prediction applicable to entire classes of allosteric modulators. These findings address a fundamental limitation in current AI-based structure prediction tools and provide a practical framework for enhancing the accuracy of computational approaches in next-generation drug design, particularly for the development of novel chemical entities that engage targets through unexplored allosteric mechanisms essential for overcoming drug resistance.

**Figure legends**:

**Figure 1. Comparison of experimental crystal structures of EGFR co-crystallized with allosteric ligands and default AlphaFold3 predictions.** The left column shows the superimposition view of predicted and experimental structures, the middle column displays the default AF3 predicted ligand poses, and the right column presents the ground truth ligand positions from crystal structures. This comparison demonstrates that default AF3 predictions produce completely incorrect prediction outcomes compared to the experimental ground truth structures.

**Figure 2. Sequence purification of EGFR to active and inactive states. A.** Structural visualization and illustration of the active state (PDB 2ITP, in blue) and inactive state (PDB 2GS7, in purple), with superimposition view highlighting the signature dynamics of the αC helix and activation loop two-turn helices; **B.** Distribution of structures predicted during M-fold sampling. The top panel is color-coded by the global pLDDT of the protein, while the lower panel is color-coded by the local pLDDT of the αC helix and activation loop two-turn helical region. RMSD with respect to both states was calculated only for the αC helix and two-turn helices region (residues 756–769 and 857–863). **C.** Distribution of RMSD with respect to the active and inactive states and voting results. The middle panel indicates normal voting, where no threshold was enforced during voting to count sequence occurrence frequency, and the bottom panel shows results with an enforced threshold of 0.15; **D.** Structure prediction results using purified sequences under different scenarios compared to prediction results using the same number of randomly selected sequences; **E.** Top ten AF2 predicted structures using sequences purified for the active state in normal voting mode and the inactive state in enforced voting mode.

**Figure 3. Comparison of AlphaFold3 prediction using purified sequences versus default settings.** AF3 prediction evaluation across three PDB crystal structure of allosteric inhibitor bound EGFR. Left column: default AF3 predictions without sequence purification. Middle column: predictions using bin 7,8 purified sequences from normal voting procedure. Right column: predictions using bin 9 purified sequences from enforced voting with threshold constraints. All predictions are color-coded according to average ligand atomic pLDDT scores to indicate prediction confidence. A total of 50 structures were generated for each experimental condition.

**Figure 4. Validation of sequence purification approach through structural comparison.** Superimposition of experimental crystal structures (light green, purple, and red) with AF3 predictions generated using purified sequences (dark green). Both protein backbone conformations and ligand binding poses exhibit remarkable agreement between predicted and experimental structures, demonstrating the effectiveness of conformational constraint through sequence purification.

**Figure 5. Conformational analysis of IL-1β allosteric binding site prediction. A.** Structural superimposition revealing that default AlphaFold3 predictions fail to capture the correct β4-5 loop conformation required for allosteric inhibitor binding, showing significant deviation from the experimental crystal structure; **B.** Comparison of β4-5 loop conformations predicted by default AF2 and AF3 models, both of which adopt conformations incompatible with cryptic pocket formation, demonstrating the limitation of standard MSA inputs in predicting allosteric conformational states; **C.** Distribution analysis of structures generated during M-fold sampling, plotted according to RMSD with respect to the β4-5 loop and β7-8 loop regions relative to the experimentally determined allosteric conformation.

**Figure 6. Performance validation of sequence purification approach for IL-1β allosteric inhibitor prediction.** Left: Comparison between experimental crystal structures and default AF3 predictions, demonstrating complete prediction failure with incorrect protein and ligand conformations. Middle: AF3 predictions generated using MSAs compiled from the top 10 most frequently occurring sequences from purified subsets, showing significant improvement in structural accuracy. Right: AF3 predictions using MSAs compiled from the top 20 most frequent sequences, achieving near-perfect alignment with experimental structures.

# Figure 1

| Superimposition | AF3 Default prediction | Ground truth |
|---|---|---|
| 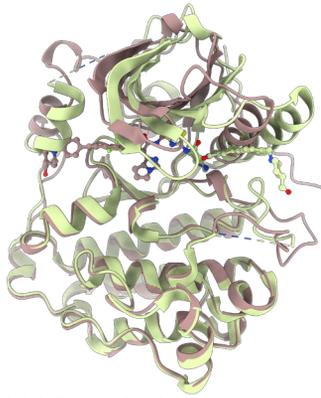 | 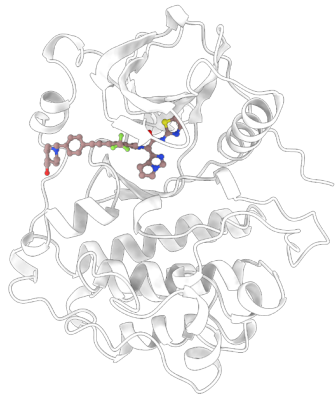 | 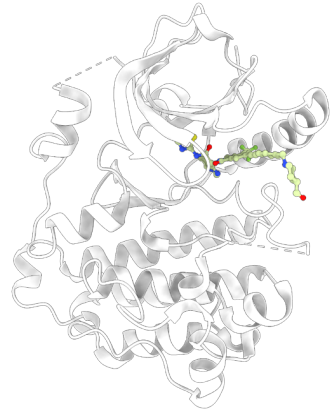 |

PDB Code: 8A2A
Ligand RMSD (pred vs true): 19.1 Å
A-loop/αC helix RMSD: 7.2 Å

| | | |
|---|---|---|
| 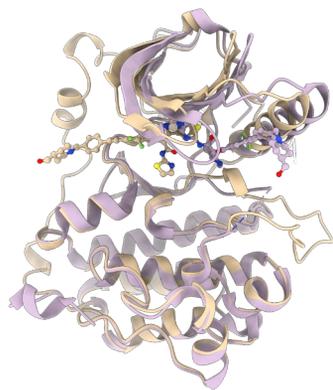 | 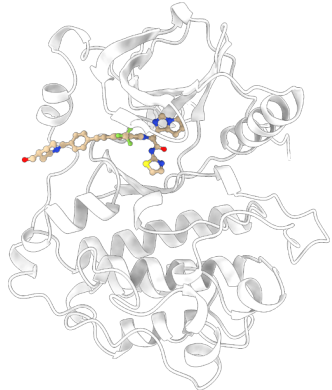 | 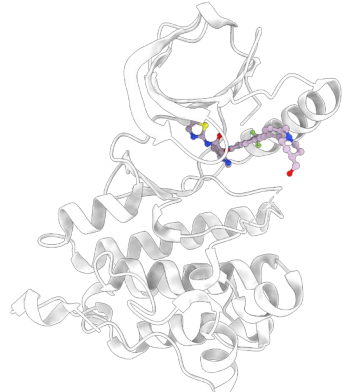 |

PDB Code: 8A2B
Ligand RMSD (pred vs true): 18.9 Å
A-loop/αC helix RMSD: 3.2 Å

| | | |
|---|---|---|
| 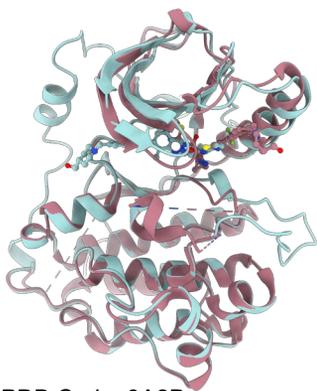 | 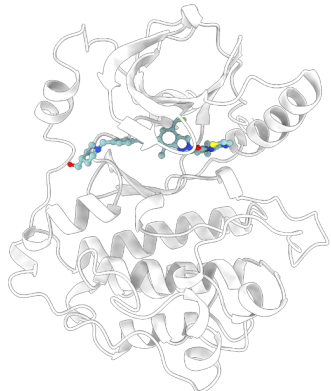 | 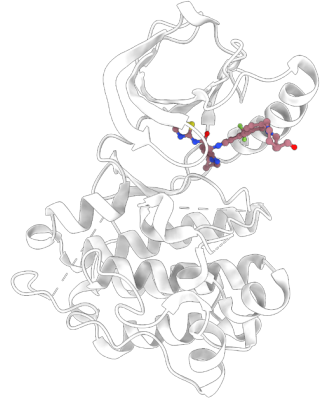 |

PDB Code: 8A2D
Ligand RMSD (pred vs true): 14.9 Å
A-loop/αC helix RMSD: 7.2 Å

**Figure 2**

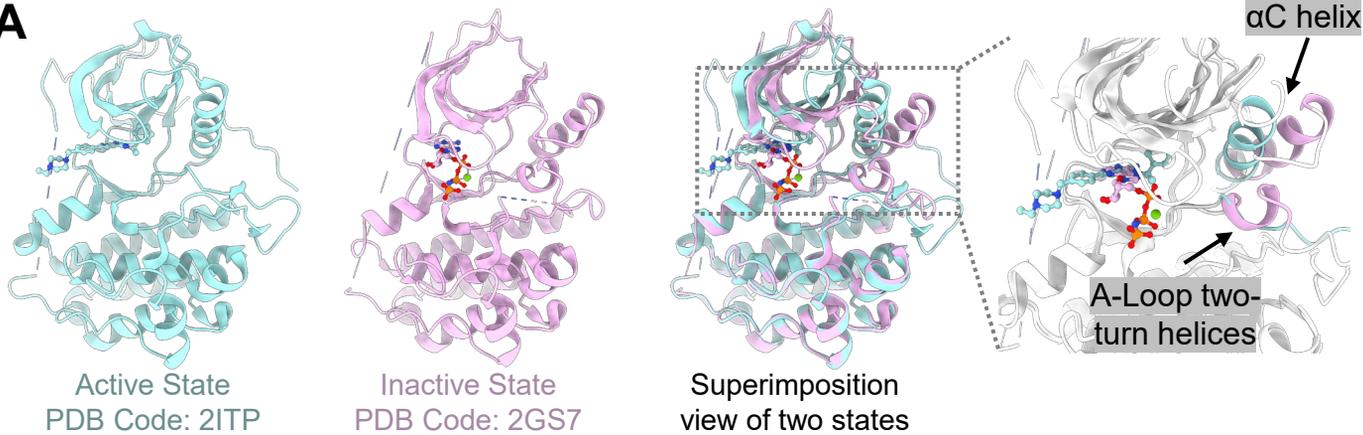
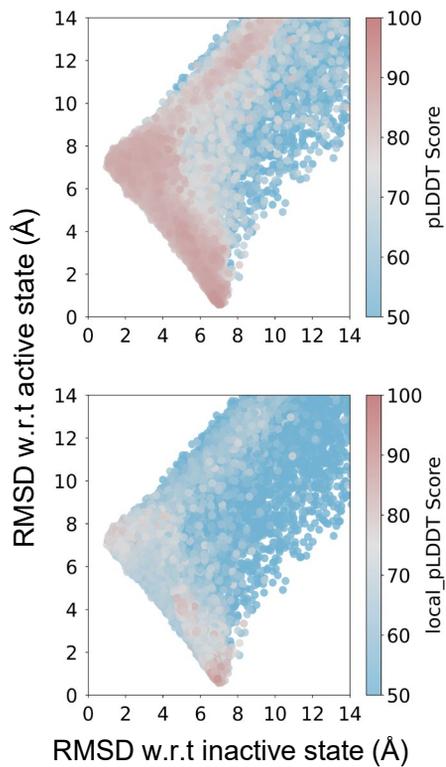
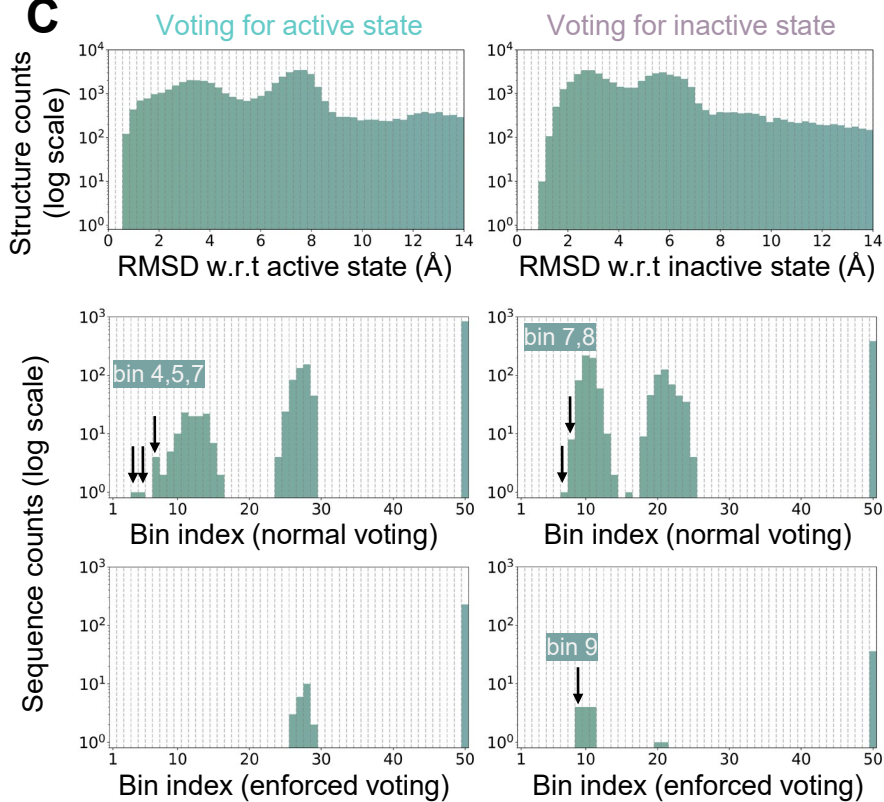
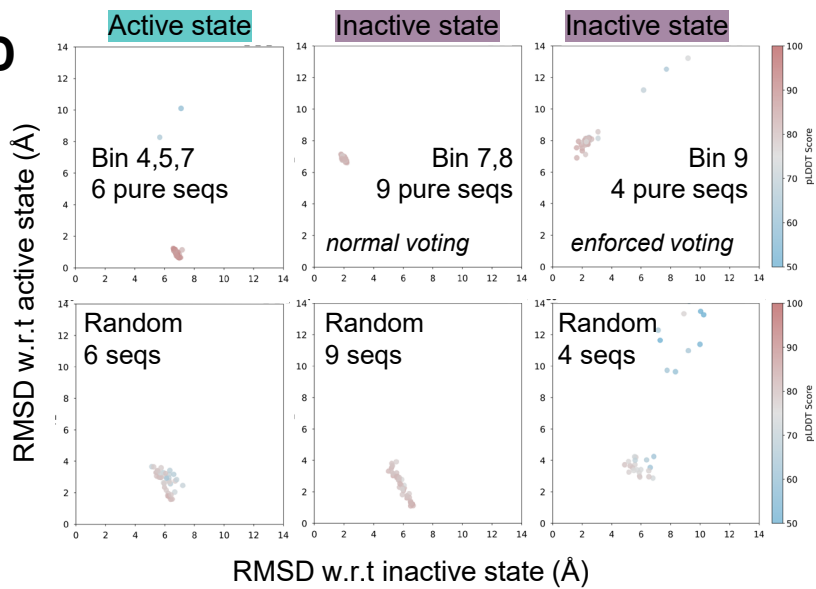
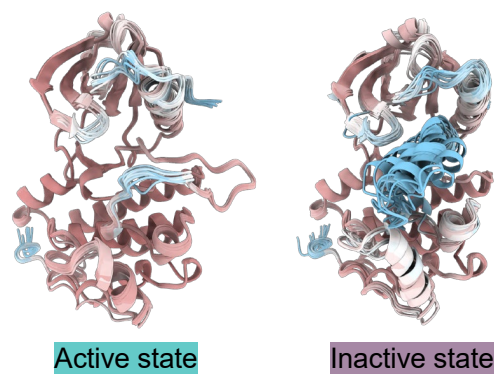

# Figure 3

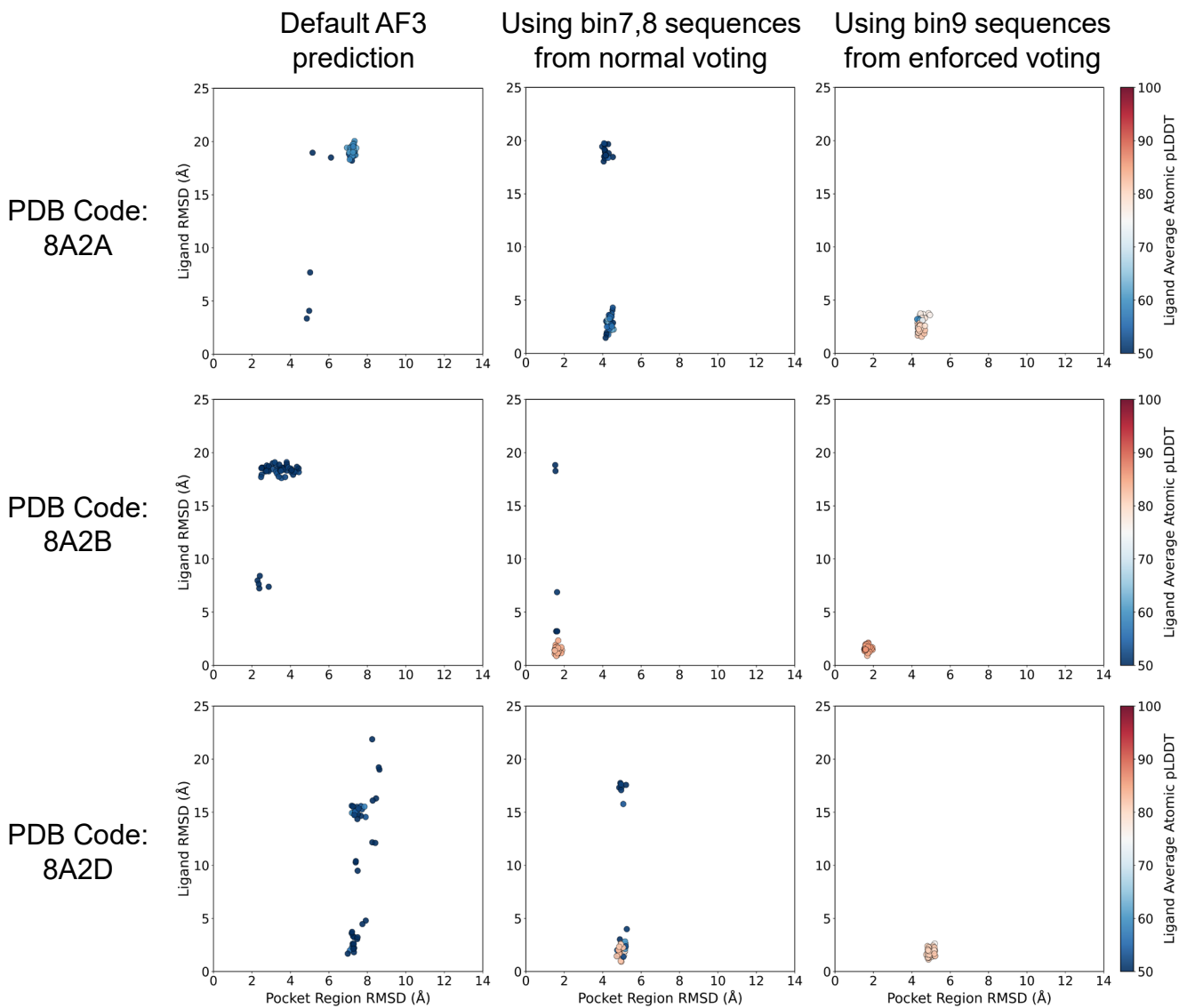

**Figure 4**

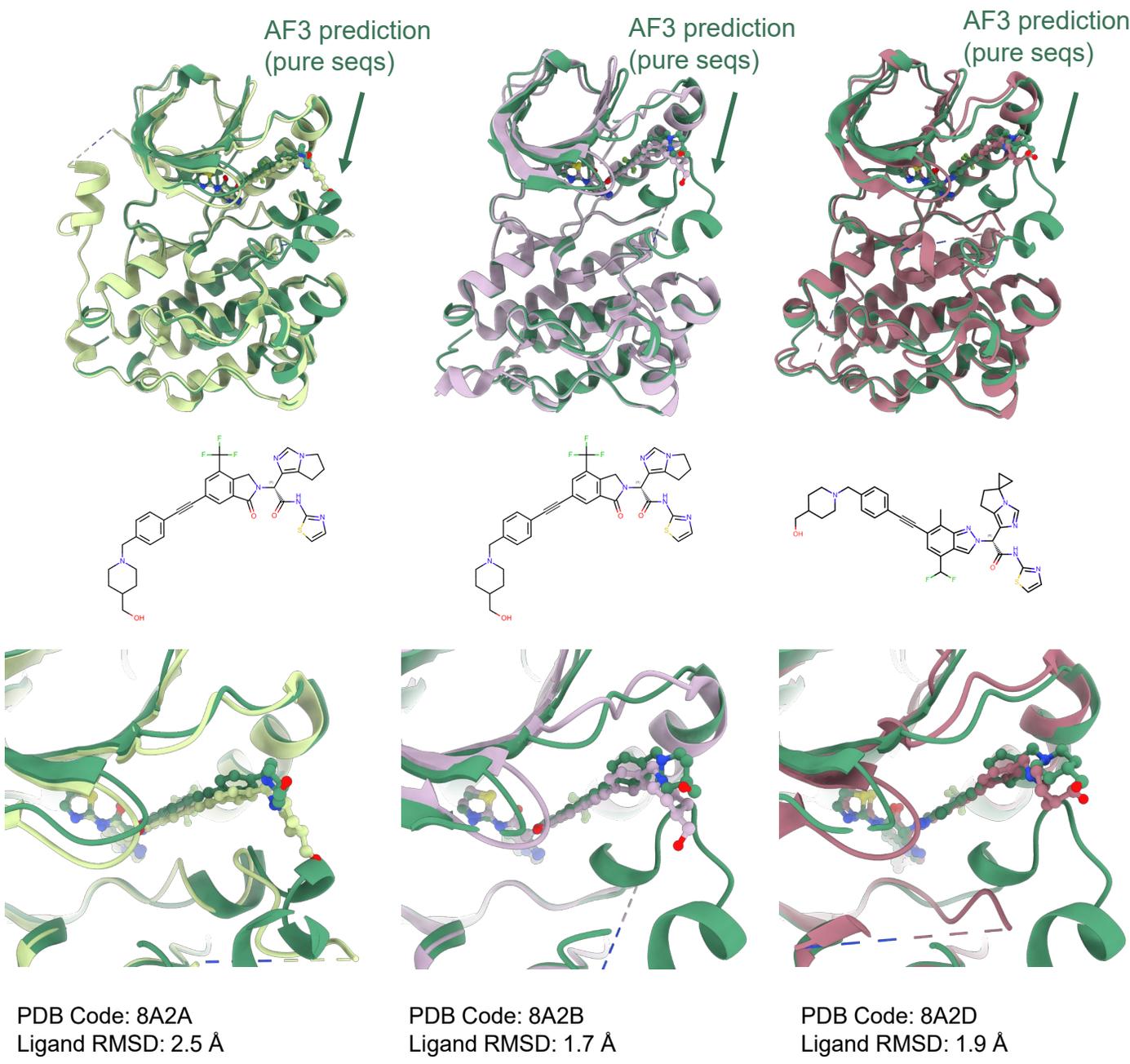

PDB Code: 8A2A
Ligand RMSD: 2.5 Å

PDB Code: 8A2B
Ligand RMSD: 1.7 Å

PDB Code: 8A2D
Ligand RMSD: 1.9 Å

# Figure 5

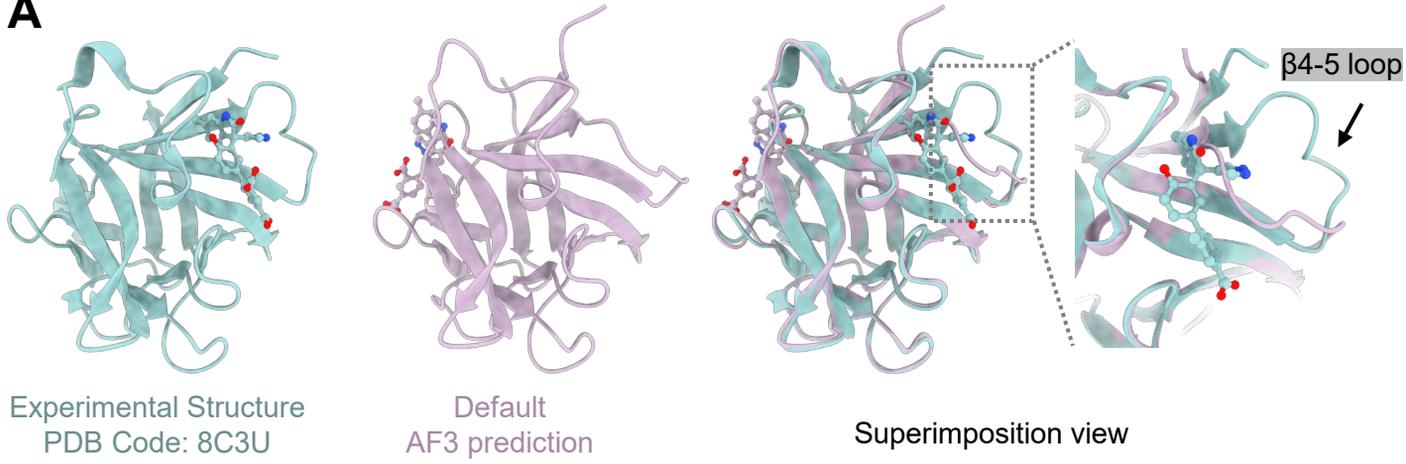

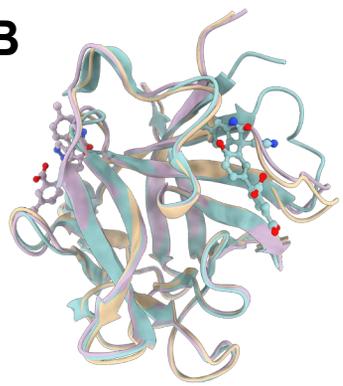

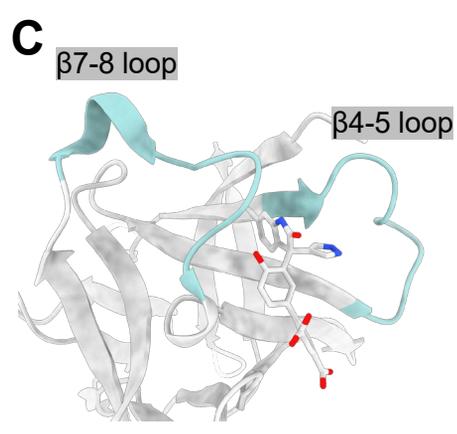

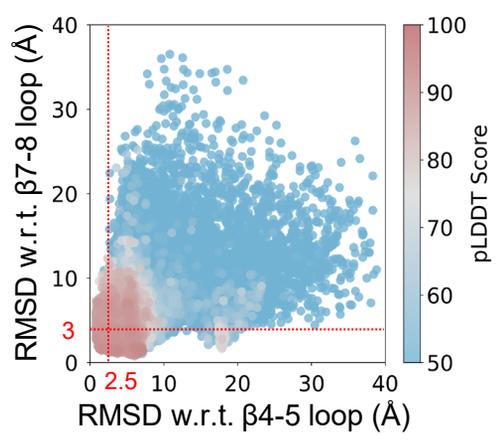

# Figure 6

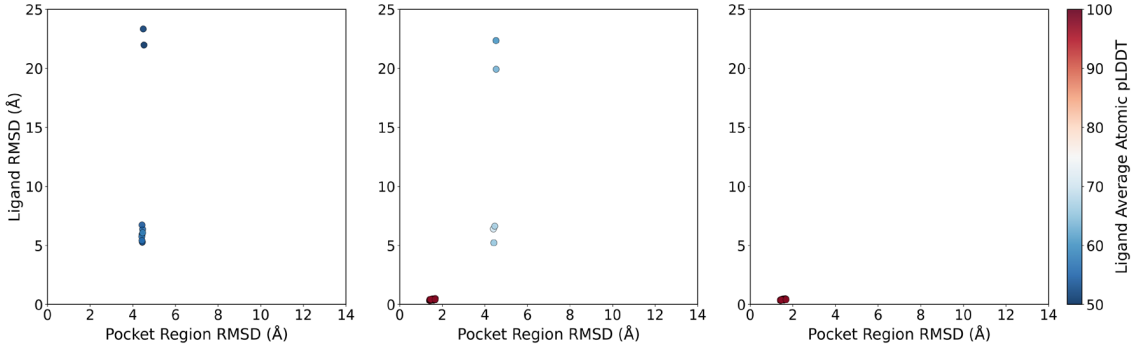

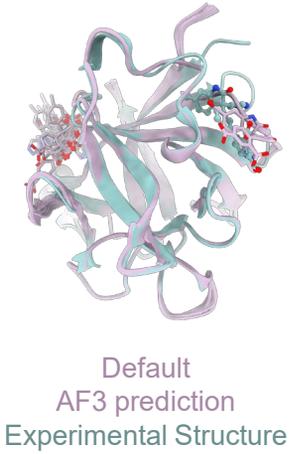

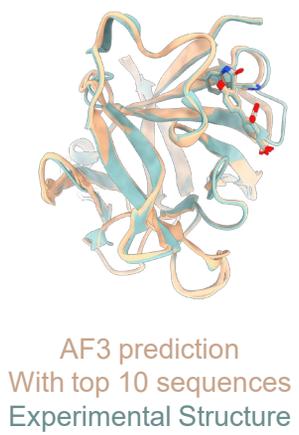

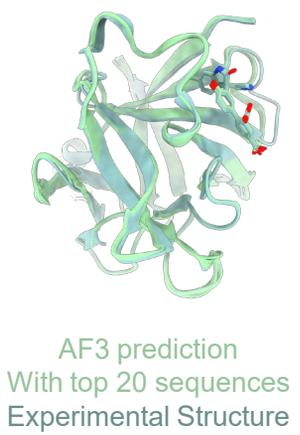

Default
AF3 prediction
Experimental Structure

AF3 prediction
With top 10 sequences
Experimental Structure

AF3 prediction
With top 20 sequences
Experimental Structure